\documentclass[12pt]{article}
\title{\bf Classical antiparticles, quantum supersymmetry anomaly
and constituent models}
\author{Yu.S.Kalashnikova\thanks{e-mail: yulia@vxitep.itep.ru},
A.V.Nefediev\thanks{e-mail: nefediev@vxitep.itep.ru}}
\date{\it Institute of Theoretical and Experimental Physics, 117218, Moscow,
Russia}
\newcommand{\be}{\begin{equation}}
\newcommand{\ee}{\end{equation}}
\newcommand{\ds}{\displaystyle}
\newcommand{\vr}{\vec{r}}
\newcommand{\vp}{\vec{p}}
\newcommand{\dvr}{\dot{\vec{r}}}
\begin{document}
\maketitle

\begin{abstract}
The two ways of constrained systems quantization are considered
from the point of view of their self-consistency at the quantum
level. With a transparent example of a particle in the external
electromagnetic field we demonstrate that the procedure of gauge
fixing turns out rather dangerous and may lead to a quantum anomaly
in the operator algebra. We discuss additional classical symmetries
as an essential element for tracing out this anomaly. The two cases
of a spinning and a spinless particles in the external
electromagnetic field are discussed to illustrate the situation.
\end{abstract}

Various classical and quantum mechanical aspects of Dirac equation
continue attracting a lot of attention. One of the main reasons for it
is that Dirac equation actually initiated the discussion of \lq\lq
negative energy states" which, in turn, gave rise to the concept of
antiparticles. While the crucial role of antiparticles in quantum
field theory cannot be overestimated, the relativistic mechanics of
antiparticles is not understood well enough (for a brief review of
the issue see \cite{1}).

The problem is rooted in the ambiguity which one encounters
describing the motion of relativistic particle in a
Lorentz--invariant way: the measure of the length along the
particle worldline can be defined only up to the sign, so that for
the particle at rest one finds
\be
d\tau=\pm dt,
\label{1}
\ee
where $d\tau$ is the infinitesimal interval of the proper time, and
$dt$ is that in the given reference frame. These both solutions are
equally valid, and it was realised many years ago \cite{2} that the
upper sign in (\ref{1}) describes the \lq\lq particle motion",
whereas the lower one corresponds to the \lq\lq motion of
antiparticle".

The possibility of classical antiparticle motion is closely
connected to the fundamental symmetries of the theory, as it was
discussed in \cite{1}, and, in a more formal way, in \cite{3}. It
was shown in \cite{3} that it was possible to perform the canonical
quantization for a free particle in a manner which allows to remove
the above--mentioned sign ambiguity and to describe the particle
and the antipaticle together within the same theory both at the
classical and quantum level.  The procedure is easily generalized
to the case of spinning Dirac particle, and ends up with the Dirac
equation in the Foldy--Wouthuysen representation \cite{4}. The
latter representation not only decouples completely the positive
and negative energy states, but also provides the Newton--Wigner
position and spin operators \cite{5} which, in contrast to the ones
of the Dirac--Pauli representation, correspond to their classical
counterparts and have clear physical meaning. The natural question,
already asked in \cite{1}, is if it is worth bothering with the Dirac
equation in any other representation? The answer is quite obvious:
what one actually needs is a theory of interacting particles.
Indeed, while there is no classical force which causes the
particle--antiparticle mixing, it easily occurs at the quantum
level. The aim of the present paper is to put this statement onto
formal grounds. Namely, we demonstrate with a simple example that
not all symmetries enjoyed by the classical spinning particle
survive at the quantum level, if the quantization is performed in a
manner which distinguishes between two signs in equation (\ref{1}).

We start with the action describing the motion of a spinning
particle in the external electromagnetic field $A_{\mu}$ (see {\it
e.g.} \cite{6}):
\be
S=\int_{\tau_i}^{\tau_f}\;Ld\tau,
\label{2}
\ee
\be
L=-\frac{\mu}{2}\dot{x}_{\nu}(\dot{x}_{\nu}-i\chi\psi_{\nu})-
\frac{i}{2}\psi_{\nu}\dot{\psi}_{\nu}-\frac{m^2}{2\mu}+\frac{i}{2}
(\psi_5\dot{\psi}_5+m\chi\psi_5)-gA_{\mu}\dot{x}_{\mu}+
\frac{ig}{2\mu}\psi_{\mu}\psi_{\nu}F_{\mu\nu}.
\label{3}
\ee

Here $\tau$ is the proper time, $x_{\mu}$ and $\psi_{\mu}$
($\mu=0,1,2,3$) are the position and Grassmannian spin variables,
the fifth Grassmannian variable $\psi_5$ is introduced to consider
massive particle, the dot means the derivative with respect to the
proper time, and $\mu$ and $\chi$ are the einbein fields, $\mu$
being a commuting and $\chi$ an anticommuting variable.

Action (\ref{2},\ref{3}) is invariant under reparametrization group
transformations
\be
\tau\to f(\tau),\quad \mu\to\frac{\mu}{\dot{f}(\tau)},\quad 
\chi\to\frac{\chi}{\dot{f}(\tau)},
\label{4}
\ee
as well as under supergauge transformations generated (in the
infinitesimal form) by the anticommuting quantity $\alpha(\tau)$:
\be
\begin{array}{l}
\delta x_{\nu}=i\alpha\psi_{\nu},\quad \delta\psi_{\nu}=-\alpha
\mu(\dot{x}_{\nu}-\frac{i}{2}\chi\psi_{\nu})\\
{}\\
\delta\mu=i\alpha\mu^2\chi,\quad \delta\chi=2\dot{\alpha},\quad
\delta\psi_5=m\alpha.
\end{array}
\label{5}
\ee

The presence of these invariances indicates that, in accordance
with Dirac \cite{7}, there should be two primary first class
constraints among the whole set of constraints for theory
(\ref{2},\ref{3}). Let us first briefly outline the standard way of
dealing with such a situation.

The conjugated momenta are defined as 
$$
p_{\nu}=\frac{\partial L}{\partial\dot{x}_{\nu}}=-\mu(\dot{x}_{\nu}-
\frac{i}{2}\chi\psi_{\nu})-gA_{\nu},
$$
\be
p_{\psi_{\nu}}=\frac{\partial
L}{\partial\dot{\psi}_{\nu}}=\frac{i}{2}\psi_{\nu},\quad 
p_{\psi_5}=\frac{\partial
L}{\partial\dot{\psi}_5}=-\frac{i}{2}\psi_5,
\label{6}
\ee
$$
\pi=\frac{\partial L}{\partial\dot{\mu}}=0,\quad
\pi_{\chi}=\frac{\partial L}{\partial\dot{\chi}}=0.
$$

The constraints invoking momenta $p_{\psi_{\mu}}$ and $p_{\psi_5}$ are of
the second class, and these variables are eliminated from the theory with Dirac
bracket
$$
\{AB\}'=
\frac{\ds\partial A}{\ds\partial x_{\nu}}\frac{\ds\partial B}
{\ds\partial p_{\nu}}-
\frac{\ds\partial A}{\ds\partial p_{\nu}}\frac{\ds\partial B}
{\ds\partial x_{\nu}}+
\frac{\ds\partial A}{\ds\partial \mu}\frac{\ds\partial B}
{\ds\partial \pi}-
\frac{\ds\partial A}
{\ds\partial \pi}\frac{\ds\partial B}{\ds\partial \mu}-
$$
\be
-A\overleftarrow{\frac{\ds\partial}{\ds\partial \chi}}
\overrightarrow{\frac{\ds\partial}{\ds\partial \pi_{\chi}}}B
-A\overleftarrow{\frac{\ds\partial}{\ds\partial \pi_{\chi}}}
\overrightarrow{\frac{\ds\partial}{\ds\partial \chi}}B
+iA\overleftarrow{\frac{\ds\partial}{\ds\partial \psi_{\mu}}}
\overrightarrow{\frac{\ds\partial}{\ds\partial \psi_{\mu}}}B-
iA\overleftarrow{\frac{\ds\partial}{\ds\partial \psi_5}}
\overrightarrow{\frac{\ds\partial}{\ds\partial \psi_5}}B
\ee

Primary constraints $\Phi_1=\pi\approx 0$ and 
$\Phi_2=\pi_{\chi}\approx 0$ give rise to the secondary ones
\be
\Phi_3=\{\Phi_1H\}'=\frac{1}{2\mu^2}\left((p+gA)^2-m^2+ig\psi_{\mu}\psi_{\nu}
F_{\mu\nu}\right),                
\label{8}
\ee                 
\be
\Phi_4=\{\Phi_2H\}'=-\frac{i}{2}\left((p+gA)\psi-m\psi_5\right),                
\label{9}
\ee                                
where Hamiltonian $H$ is given by the expression
\be
H=-\frac{\ds 1}{\ds 2\mu}[(p+gA)^2-m^2+igF_{\mu\nu}\psi_{\mu}\psi_{\nu}]
+\frac{\ds i}{\ds 2}\chi[(p+gA)\psi-m\psi_5]
+\lambda_{\chi}\pi_{\chi}+\lambda\pi,
\label{10}
\ee
with primary constraints $\Phi_1=\pi$ and $\Phi_2=\pi_{\chi}$  
added with Lagrange multipliers $\lambda$ and 
$\lambda_{\chi}$.

The only nonzero brackets of the constraints 
\be
\begin{array}{l}
\{\Phi_1\Phi_3\}'=-\frac{\ds 1}{\ds 2\mu^3}\Phi_3,\\
{}\\
\{\Phi_4\Phi_4\}'=i\Phi_3,
\end{array}
\label{11}
\ee
vanish at the constraint surface together with the Hamiltonian, 
so that the set of constraints $\{\Phi_i\}$, $i=1,\;2,\;3,\;4$, is the first
class one.

The way to quantize a theory in the presence of the first class constraints was
suggested by Dirac. In the given case the quantization {\it a la} Dirac is
performed setting 
\be
\hat{p}_{\mu}=i\frac{\partial}{\partial x_{\mu}},\quad
\psi_{\mu}=\frac{1}{\sqrt{2}}\gamma_5\gamma_{\mu},\quad 
\psi_5=\frac{1}{\sqrt{2}}\gamma_5.
\label{12}
\ee

The physically relevant constraints $\Phi_3$ and $\Phi_4$ provide the 
following equations for the wave function:
\be
\hat{\Phi}_{KG}\Psi=\left((p+gA)^2-m^2-\frac{i}{2}g\sigma_{\mu\nu}F_{\mu\nu}\right)\Psi=0,                
\label{13}
\ee                 
\be
\hat{\Phi}_D\Psi=\gamma_5\left(\gamma (p+gA)-m\right)\Psi=0,                
\label{14}
\ee
yielding Hamiltonian in the Dirac--Pauli representation
\be
\hat{H}=\vec{\alpha}(\vec{p}-g\vec{A})+\gamma_0 m +gA_0.
\label{15}
\ee

The most important point here is that algebra of constraints 
(\ref{11}) remains closed at the quantum level with operator
realisation 
(\ref{12}):
\be
[\hat{\Phi}_{KG},\hat{\Phi}_{KG}]_-=0,\quad
[\hat{\Phi}_{KG},\hat{\Phi}_{D}]_-=0,\quad [\hat{\Phi}_D,\hat{\Phi}_D]_+=-
\hat{\Phi}_{KG},
\label{16}
\ee
that makes equations (\ref{13}) and (\ref{14}) for the wave function compatible
with one another. 

Now we consider an alternative way of dealing with the constrained theory. One
can impose additional constraints which fix the gauges; with these extra constraints
the degeneracy of the theory is removed, all the constraints become the second class
ones, and the resulting theory yields a Hamiltonian which is nonzero at the
constraint surface. The consistent procedure for theory (\ref{2},\ref{3}) is
described in \cite{8}; for our purposes it is enough to present a simplified version.
To this end we fix only one gauge in reparametrization group (\ref{4}) setting
\be
x_0=\tau
\label{17}
\ee
in Lagrangian (\ref{3}) and anticipating the quantization at the time-like
hyper-surface. In such a way the upper sign in equation (\ref{1}) is chosen from the
very beginning.

In what follows we shall consider the stationary problem with the four--potential
$A_{\mu}(x_0,\vec{r})$ depending only on $\vr$.

With gauge fixing condition (\ref{17}) the Lagrangian takes the form
$$
L=-\frac{\mu}{2}+\frac{\mu\dvr^2}{2}+\frac{i}{2}\mu\chi\psi_0-
\frac{i}{2}\mu\chi(\dvr\vec{\psi})-
\frac{i}{2}\psi_{\nu}\dot{\psi}_{\nu}
-\frac{m^2}{2\mu}+
$$
\be
+\frac{i}{2}(\psi_5\dot{\psi}_5+m\chi\psi_5)-gA_0+g\vec{A}\dvr+
\frac{ig}{2\mu}\psi_{\mu}\psi_{\nu}F_{\mu\nu},
\label{18}
\ee
and the conjugated momenta are
$$
\vp=\frac{\partial L}{\partial\dvr}=\mu(\dvr-\frac{i}{2}\chi\vec{\psi})+g\vec{A},
$$
\be
p_{\psi_{\nu}}=\frac{\partial
L}{\partial\dot{\psi}_{\nu}}=\frac{i}{2}\psi_{\nu},\quad 
p_{\psi_5}=\frac{\partial
L}{\partial\dot{\psi}_5}=-\frac{i}{2}\psi_5,
\label{19}
\ee
$$
\pi=\frac{\partial L}{\partial\dot{\mu}}=0,\quad
\pi_{\chi}=\frac{\partial L}{\partial\dot{\chi}}=0.
$$

Again we eliminate redundant variables 
$p_{\psi_{\mu}},\;p_{\psi_5}$
defining the Dirac brackets as
$$
\{AB\}'=
\frac{\ds\partial A}{\ds\partial \vec{p}}\frac{\ds\partial B}
{\ds\partial \vec{x}}-
\frac{\ds\partial A}{\ds\partial \vec{x}}\frac{\ds\partial B}
{\ds\partial \vec{p}}-
\frac{\ds\partial A}{\ds\partial \pi}\frac{\ds\partial B}{\ds\partial \mu}+
\frac{\ds\partial A}{\ds\partial \mu}\frac{\ds\partial B}{\ds\partial \pi}-
$$
\be
-A\overleftarrow{\frac{\ds\partial}{\ds\partial \chi}}
\overrightarrow{\frac{\ds\partial}{\ds\partial \pi_{\chi}}}B
-A\overleftarrow{\frac{\ds\partial}{\ds\partial \pi_{\chi}}}
\overrightarrow{\frac{\ds\partial}{\ds\partial \chi}}B
+iA\overleftarrow{\frac{\ds\partial}{\ds\partial \psi_0}}
\overrightarrow{\frac{\ds\partial}{\ds\partial \psi_0}}B
-iA\overleftarrow{\frac{\ds\partial}{\ds\partial \vec{\psi}}}
\overrightarrow{\frac{\ds\partial}{\ds\partial \vec{\psi}}}B-
iA\overleftarrow{\frac{\ds\partial}{\ds\partial \psi_5}}
\overrightarrow{\frac{\ds\partial}{\ds\partial \psi_5}}B.
\label{20}
\ee

The Hamiltonian takes the form 
\be
\begin{array}{l}
H=\frac{\ds \mu}{\ds 2}+\frac{\ds (\vec{p}-g\vec{A})^2+m^2}{\ds 2\mu}
-\frac{\ds i}{\ds 2}\chi(\mu\psi_0-(\vec{p}-g\vec{A})\vec{\psi}+m\psi_5)\\
{}\\
+\frac{\ds ig}{\ds \mu}F_{0i}\psi_0\psi_i
-\frac{\ds ig}{\ds 2\mu}F_{ik}\psi_i\psi_k+gA_0,
\end{array}
\label{21}
\ee
and the remaining primary constraints
\be
\begin{array}{l}
\varphi_1=\pi\\
\varphi_2=\pi_{\chi}
\end{array}
\label{22}
\ee
give rise to to the secondary constraints

\be
\begin{array}{l}
\varphi_3=\{\varphi_1H\}'=
-\frac{\ds 1}{\ds 2}+\frac{\ds (\vec{p}-g\vec{A})^2+m^2}{\ds 2\mu^2}
+\frac{\ds i}{\ds 2}\chi\psi_0
+\frac{\ds ig}{\ds \mu^2}F_{0i}\psi_0\psi_i
-\frac{\ds ig}{\ds 2\mu^2}F_{ik}\psi_i\psi_k,\\
{}\\
\varphi_4=\{\varphi_2H\}'=\frac{\ds i}{\ds 2}
(\mu\psi_0-(\vec{p}-g\vec{A})\vec{\psi}+m\psi_5).
\end{array}
\label{23}
\ee

Condition (\ref{17}) fixes only the gauge in the reparametrization group, and
Lagrangian (\ref{18}) is still invariant under supergauge transformations. This
means that constraint matrix $C_{ij}=\{\varphi_i\varphi_j\}'$ is still degenerate. We
demonstrate it explicitly introducing the modified brackets
\be
\{AB\}^*=\{AB\}'-\{A\varphi_1\}'C_{13}^{-1}\{\varphi_3B\}'
-\{A\varphi_3\}'C_{31}^{-1}\{\varphi_1B\}',
\label{24}
\ee
where 
\be
C_{13}=\{\varphi_1\varphi_3\}'=
\frac{\ds (\vec{p}-g\vec{A})^2+m^2}{\ds 2\mu^3}
+\frac{\ds 2ig}{\ds \mu^3}F_{0i}\psi_0\psi_i
-\frac{\ds ig}{\ds \mu^3}F_{ik}\psi_i\psi_k.
\label{25}
\ee

Then the explicit calculation shows that the odd pair of constraints is of the first
class,
\be
\{\varphi_2\varphi_2\}^*=\{\varphi_2\varphi_4\}^*=0,
\label{26}
\ee
\be
\{\varphi_4\varphi_4\}^*=0,
\label{27}
\ee
and the physically relevant constraint $\varphi_4$ commutes with the Hamiltonian:
\be
\{\varphi_4H\}^*=0.
\label{28}
\ee

The physical Hamiltonian 
\be
H_{ph}=\mu_0+gA_0+\frac{ig}{\mu_0}F_{0i}\psi_0\psi_i
\label{29}
\ee
and the physical Dirac constraint
\be
\varphi_D=\mu_0\psi_0-(\vec{p}-g\vec{A})\vec{\psi}+m\psi_5
\label{30}
\ee
are obtained on substituting the solution 
\be
\mu_0=\sqrt{(\vec{p}-g\vec{A})^2+m^2-igF_{ik}\psi_i\psi_k}
\label{31}
\ee
of the constraint equation $\varphi_3=0$
\footnote{The solution for $\mu_0$ does not contain part proportional to
$\psi_0$ as it is left explicitly in Hamiltonian (\ref{29}), whereas in 
Dirac constraint (\ref{30}) it would vanish due to the Grassmann nature of 
$\psi_0$.}. 
Note that to arrive at algebra
(\ref{27}), (\ref{28}) as well as at forms (\ref{29}), (\ref{30}) it is necessary to
take into account the relations 
\be
\psi_{\mu}\psi_{\nu}=-\psi_{\nu}\psi_{\mu}, \quad
\psi_{5}\psi_{\mu}=-\psi_{\mu}\psi_{5},\quad \psi_5\psi_5=0
\label{32}
\ee
for the elements of the Grasmannian algebra.

The theory should be quantized with bracket (\ref{24}) in the usual way, setting
$\hat{\vec{p}}=-i\frac{\partial}{\partial \vr}$ and 
$\psi_{\mu}=\frac{1}{\sqrt{2}}\gamma_5\gamma_{\mu}$, 
$\psi_5=\frac{1}{\sqrt{2}}\gamma_5$. The wave function should not only satisfy the
Schr{\" o}dinger equation
\be
\hat{H}_{ph}\Psi=\left(\hat{\mu}_0+gA_0+\frac{ig}{2\hat{\mu}_0}F_{0i}\sigma_{0i}
\right)\Psi=E\Psi,
\label{33}
\ee
but also the constraint equation
\be
\hat{\varphi}_D\Psi=\left(\hat{\mu}_0\gamma_0-(\vec{p}-g\vec{A})\vec{\gamma}-
m\right)\Psi=0.
\label{34}
\ee

It is easy to see, however, that the quantum algebra of the Hamiltonian and the Dirac
constraint is not closed,
\be
[\hat{H}_{ph}\hat{\varphi}_D]_-\neq 0,
\label{35}
\ee
so equations (\ref{33}) and (\ref{34}) are not compatible. We stress that it is not a
problem of the operator ordering but it takes place because there is no relation for
$\gamma$-matrices similar to (\ref{32}) for Grassmannian variables.

There are, of course, special types of the external field configurations for which
the Hamiltonian commutes with the Dirac constraint at the quantum level
\footnote{Note that it is the electric field
to be responsible for self-inconsistency (\ref{35}).}
(see {\it
e.g.} \cite{9}) but for the general case one has encountered a quantum supersymmetric
anomaly which affects the physical results. In particular, the well--known Darwin 
term in the Hamiltonian is completely lost with such a kind of gauge fixing.

One can go further, and fix the gauge in the supergauge group (\ref{5}) too, as it
was done in \cite{8}. Nevertheless, the resulting quantum theory \cite{9} has not got
the Darwin term restored. Moreover, as with the complete gauge fixing all the
constraints are already of the second class, there is no additional equation for the
wave function like (\ref{34}), and one should not impose extra compatibility
requirements. 

We can see now that one is very lucky to be able to pin-point the source of troubles
with spinning particle. Indeed, let us consider the case of scalar particle, where
there are no spin variables and the only symmetry is the reparametrization one.
Skipping the details we write out the ultimate Klein--Gordon equation for the wave
function
\be
((\hat{p}+gA)^2-m^2)\Psi=0
\label{36}
\ee
for the case of no gauge fixing procedure {\it a la} Dirac, and the Schr{\" o}dinger
equation
\be
\hat{H}_{ph}\Psi=\left(\sqrt{m^2+(\hat{\vec{p}}-e\vec{A})^2}+gA_0\right)\Psi=E\Psi
\label{37}
\ee 
for the case of gauge fixed by condition (\ref{17}). Equations (\ref{36}) and
(\ref{37}) yield different spectra, and the Darwin term, which also exists for the
scalar particle (see {\it e.g.} \cite{10}) is lost again in (\ref{37}). In contrast to the Dirac particle case,
there is no extra symmetry and no way to find out how it could happen.

The anomaly discussed is not an artifact of the time-like gauge fixing (\ref{17});
the classical antiparticles do exist under any assumption on the evolution parameter
$\tau$. The gauge conditions which forbid the particle--antiparticle mixing at the
quantum level exclude some physical phase space trajectories and are not admissible.
For the time-like gauge fixing one truncates the phase space excluding the negative
energy states, but, for example, for another popular light-cone gauge fixing the point
$p_+=0$ is excluded from the phase space.

While the situation is rather trivial in the transparent case of external field
discussed above, in a more complicated cases of interacting particles one meets even
more confusions. For example, the exactly solvable problem of a quark--antiquark pair
in 1+1 space-time interacting via string was considered \cite{11} with two different
versions of the reparametrization group gauge fixing. The quantization in the
proper-time and the light-cone  gauges was performed yielding different quantum
spectra. On the other hand, the only symmetry group for this theory at our disposal
is the Poincar{\' e} one, and in both gauges the quantum Poincar{\' e} algebra
appears to be closed.

Do our findings mean that as far as the uncontrolled deficiencies take place, one
has to completely abandon the first quantization procedure? The answer is,
of course,
\lq\lq no". The quantization {\it a la} Dirac, when the first class constraints are
left in peace, is safe. One may develop the first--quantized field theory from the
Feynman--Schwinger representation approach nicely reproducing the Feynman rules
\cite{12}. The technical simplicity and physical transparency of such a path integral
formulation is obvious, as well as the advantage of being back to basic quantum
mechanics.

We conclude with some phenomenological implications. The real particle--antiparticle
problem difficulties start with the important case of QCD, where in the absence of
exact solutions one relies upon models, all of which involving linearly rising force
potentially dangerous from the point of view of the 
Klein paradox. The latter observation
leads to the belief \cite{13} that the proper quantum mechanical reduction of the
underlying field theory should include the \lq\lq no-pair" assumption. We do not
share this belief as there is no way within the field theory to imply such an
assumption in a self-consistent way. The phenomenological successes of constituent
quark models tell us that the quark backward motion is suppressed, but this
suppression should be dynamical one rather than imposed by hand--waving arguments. We
are not able to prove this statement and refer to the example of 1+1 't Hooft model
\cite{14}, where confinement does occur whereas the spectrum is conveniently bound
from below without {\it ad hoc} \lq\lq no-pair" assumption. Besides, numerical solutions for
the 't Hooft model in the mesonic rest frame exist \cite{15}, which explicitly exhibit
the backward motion suppression. Moreover, the motion of a quark in the field of a
static antiquark source was considered in this model \cite{16}, and the quark
Hamiltonian was obtained both in Dirac--Pauli and Foldy--Wouthuysen representations
demonstrating explicitly that the theory prevents itself from the Klein paradox.
\bigskip

This work is supported by grants 96-02-19184a, 97-02-16404 and 96-15-96740
of Russian Fundamental Research Foundation.

\end{document}